\documentclass[10pt,twocolumn,twoside]{IEEEtran}
\usepackage{amsfonts,dsfont,amssymb,amsbsy,amsmath,paralist,theorem,bm,ifthen,color}
\usepackage[pdfstartview=FitH,bookmarksnumbered,unicode,bookmarksopen=true]{hyperref}
\usepackage{graphicx}
\usepackage{epstopdf}
\usepackage{multirow}
\usepackage{subfigure,relsize}
\usepackage{mathrsfs}
\usepackage{bm}
\usepackage{stfloats}
\usepackage{url}
\usepackage[noadjust]{cite}
\usepackage{cases,booktabs}
\usepackage{graphicx,algorithmic,algorithm,relsize}
\usepackage[justification=centering]{caption} 
\usepackage[table]{xcolor}

\newtheorem{lem}{Lemma}

\newcommand\pin{\ensuremath{{\rm Pin}}}

\newcommand\hb{\ensuremath{{\bf h}}}

\newcommand\psib{\ensuremath{{\bm \psi}}}

\newcommand\Nset  {\ensuremath{{\mathcal{N}}}}

\newcommand\st    {\ensuremath{{\rm s.t.}}}

\graphicspath{{fig/}}

\definecolor{green}{RGB}{34	195	46}
\definecolor{red}{RGB}{220 0 0}

\usepackage{graphicx} 
\allowdisplaybreaks[4]

\title{\LARGE Rate Maximization for Downlink Pinching-Antenna Systems}

\author{Yanqing Xu, \IEEEmembership{Member, IEEE,}
        Zhiguo Ding, \IEEEmembership{Fellow, IEEE,}
        and George K. Karagiannidis, \IEEEmembership{Fellow, IEEE}
        \thanks{\smaller[1] Y. Xu is with the School of Science and Engineering, The Chinese University of Hong Kong, Shenzhen, 518172, China (email: xuyanqing@cuhk.edu.cn).}
        \thanks{\smaller[1] Z. Ding was with the Department of Computer and Information Engineering, Khalifa University, Abu Dhabi, UAE, and is with the Department of Electrical and Electronic Engineering, the University of Manchester, Manchester, UK, M1 9BB.  (email: zhiguo.ding@ieee.org).}
        \thanks{\smaller[1] G. K. Karagiannidis is with the Department of Electrical and Computer Engineering, Aristotle University of Thessaloniki, 541 24 Thessaloniki, Greece (e-mail: geokarag@auth.gr). } \thanks{\smaller[1] This work has been submitted to the IEEE for possible publication.
        Copyright may be transferred without notice, after which this version may no longer be accessible.}\vspace{-3mm}
}

\date{\today}

\begin{document}

\maketitle

\begin{abstract}
In this letter, we consider a new type of flexible-antenna system, termed \textit{pinching-antenna}, where multiple low-cost pinching antennas, realized by activating small dielectric particles on a dielectric waveguide, are jointly used to serve a single-antenna user. Our goal is to maximize the downlink transmission rate by optimizing the locations of the pinching antennas. However, these locations affect both the path losses and the phase shifts of the user's effective channel gain, making the problem challenging to solve.
To address this challenge and solve the problem in a low complexity manner, a relaxed optimization problem is developed that minimizes the impact of path loss while ensuring that the received signals at the user are constructive. 
This approach leads to a two-stage algorithm: in the first stage, the locations of the pinching antennas are optimized to minimize the large-scale path loss; in the second stage, the antenna locations are refined to maximize the received signal strength. Simulation results show that pinch-antenna systems significantly outperform conventional fixed-location antenna systems, and the proposed algorithm achieves nearly the same performance as the highly complex exhaustive search-based benchmark.

\end{abstract}

\vspace{-2mm}
\begin{IEEEkeywords}
     Pinching antenna, flexible-antenna system, downlink rate maximization, line-of-sight communication.
\end{IEEEkeywords}

\vspace{-5mm}
\section{Introduction} 
Recently, flexible-antenna systems, such as fluid-antenna systems and movable-antenna systems, have been studied for their advantages in reconfiguring wireless channels \cite{wong2020fluid,zhu2023movable}. 
Unlike traditional fixed-antenna systems, flexible-antenna systems provide the ability to adjust antenna locations at the transceiver, thus improving channel conditions \cite{wong2021fluid,zhu2023movable2}.
However, in traditional flexible-antenna systems, antenna movement is limited to the wavelength scale, resulting in limited influence on large-scale path loss.
Recently, the pinching antenna has been proposed as a more promising flexible antenna system to overcome the bottlenecks in conventional flexible antenna systems \cite{suzuki2022pinching,ding2024flexible}.  Specifically, by using a dielectric waveguide as the transmission medium, pinching antennas can be repositioned through two methods: one by dynamically activating dielectric particles placed at the preconfigured locations along the waveguide, and the other by making the pinching antennas movable along a pre-installed track parallel to the waveguide. Both methods provide highly flexible antenna deployments, enabling dynamic reconfiguration of the antennas.
In addition, compared to conventional flexible antennas, the pinching-antenna system is less expensive and easier to install because the pinching mechanism involves simply adding or removing dielectric materials, making it well-suited for environments where system adaptability and cost-effectiveness are critical, such as industrial Internet of Things (IoT) or urban deployments.

\begin{figure}[!t]
	\centering
	\includegraphics[width=0.82\linewidth]{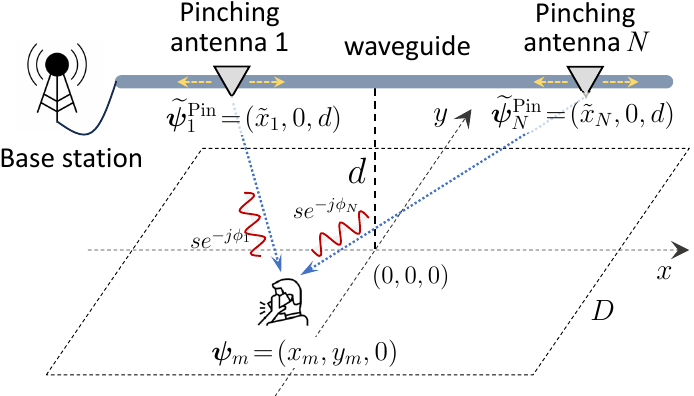}\\
        \captionsetup{justification=justified, singlelinecheck=false, font=small}	
        \caption{\small A pinching-antenna system with  $N$ pinching antennas activated on one waveguide.} \label{fig: system model} \vspace{-6mm}
\end{figure} 

In this paper, we consider a downlink pinching-antenna system. As the first study to explore the pinching-antenna system from a system design perspective, the primary goal of this work is to identify the advantages and design challenges associated with pinching-antenna systems, while offering useful insights into effective system design strategies.
To this end, we consider a system setting, where multiple pinching antennas are deployed on a waveguide to serve a single-antenna user, as shown in Fig. \ref{fig: system model}. 
The interested problem is to maximize the downlink transmission data rate by optimizing the locations of the pinching antennas. 
However, unlike fluid/movable antennas, the locations of pinching antennas influence the user's effective channel gain through two factors: path loss and dual phase shifts caused by signal propagation inside and outside the waveguide, presenting significant challenges in addressing the problem.
To tackle this challenge and solve the formulated problem in a low-complexity manner, a relaxed optimization problem that minimizes the impact of path loss while ensuring that the received signals at the user are constructive is developed.
This approach leads to a two-stage algorithm: in the first stage, the locations of pinching antennas are optimized to minimize the large-scale path loss; in the second stage, the antenna locations are refined to maximize the receive signal strength. 
The challenge for the addressed optimization problem can be reflected by the fact that the problem to be solved in the first stage is nonconvex and difficult to handle. Intriguingly, we theoretically reveal that the objective function of this problem is unimodal. 
Leveraging this revealed property, a globally optimal solution of the antenna locations is obtained in closed-form.
Then, with the obtained locations, a succinct location refinement scheme is proposed to guarantee the constructive signal receptions at the user.
Simulation results demonstrate that pinching-antenna systems significantly outperform conventional antenna systems and the proposed algorithm achieves nearly the same performance as the high complexity exhaustive search-based benchmark.

\vspace{-3mm}
\section{System Model and Problem Formulation}
Consider a downlink communication system, where a base station (BS) with $N$ antennas serves a single-antenna mobile user. The user is randomly deployed within a square area of side length $D$, as illustrated in Fig. \ref{fig: system model}. 

\vspace{-3mm} 
\subsection{Conventional Antenna System}
We first consider the conventional antenna system with fixed-position antennas, where the $N$ BS antennas are assumed to be deployed right above the centroid of the square area with a height $d$. The location of the $n$-th antenna is denoted by $\ensuremath{\bm \bar \psi}_n \!=\! [\bar x_n,0,d], \forall n \!\in\! \Nset \!\triangleq\! \{1,...,N\}$. The spacing between two neighboring antennas is set as $\Delta$ to avoid antenna coupling. As per the spherical wave channel model \cite{zhang2022beam}, the channel vector between the antennas and the user is given by \vspace{-3mm}

\begin{small}
    \begin{align} \label{eqn: h convention}
        \!\!\hb^{\rm Conv} = \bigg[ \frac{\eta^{\frac{1}{2}} e^{-j \frac{2\pi}{\lambda}\|\bm{\psi}_m - \boldsymbol{\bar\psi}_1\|}}{\|\boldsymbol{\psi}_m - \boldsymbol{\bar\psi}_1\|},..., \frac{\eta^{\frac{1}{2}} e^{-j \frac{2\pi}{\lambda}\|\boldsymbol{\psi}_m - \boldsymbol{\bar\psi}_N\|}}{\|\boldsymbol{\psi}_m - \boldsymbol{\bar\psi}_N\|}\bigg]^\top\!,
    \end{align}
\end{small}%
where $\ensuremath{\bm\psi}_m = [x_m, y_m, 0]$ represents the location of the mobile user, $\eta = \frac{c^2}{16\pi^2 f_c^2}$ is a constant where $c$ denotes the speed of light, $f_c$ is the carrier frequency, and $\lambda$ is the wavelength in free space.
The achievable data rate in the conventional antenna system is given by $R^{\rm Conv} = \log(1+ P\|\hb^{\rm Conv}\|_2^2/\sigma_w^2)$, where $P$ is the transmit power, and $\sigma_w^2$ is the power of the additive white Gaussian noise at the user.

\vspace{-3mm}
\subsection{Pinching-Antenna Systems}
The considered pinching-antenna system is shown in Fig. \ref{fig: system model}, where $N$ pinching antennas are mounted on a waveguide to jointly serve the user. The waveguide is aligned parallel to the $x$-axis at a height $d$. The transmit signal is represented by $s$, while the phase shift of the signal received from the $n$-th pinching antenna is denoted by $\phi_n, \forall n \in \Nset$, and the locations of the user and the pinching antennas are denoted by $\psib_m = [x_m, y_m,0]$ and $\widetilde\psib_n^{\pin} = [\tilde x_n, 0,d], \forall n \in \Nset$, respectively. The channel coefficient between the $n$-th pinching antenna and the user can be written as \vspace{-4mm}

\begin{small}
\begin{align} \label{eqn: channel coefficient pinching}
    \!\!h_n^{\pin} \!=\!\underbrace{\frac{\eta^{\frac{1}{2}} }{\|\boldsymbol{\psi}_m \!-\! \boldsymbol{\widetilde \psi}_n^{\pin}\|}}_{\text{free-space path loss}} \cdot \underbrace{e^{-j \frac{2\pi}{\lambda}\|\bm{\psi}_m - \boldsymbol{\widetilde \psi}_n^{\pin}\|}}_{\text{free-space phase shift}} \cdot \underbrace{e^{-j \frac{2\pi}{\lambda_g}\|\widetilde\psib^{\pin}_0 - \boldsymbol{\widetilde \psi}_n^{\pin}\|}}_{\text{in-waveguide phase shift}},
\end{align}
\end{small}%
where $\widetilde\psib_0^{\pin} = [\tilde x_0,0,d]$ denotes the location of the feed point of the waveguide, and $\lambda_g = \frac{\lambda}{n_{\rm neff}}$ denotes the guided wavelength with $n_{\rm neff}$ signifying the effective refractive index of a dielectric waveguide \cite{pozar2021microwave}. 
As observed from \eqref{eqn: channel coefficient pinching}, the channel in the pinching-antenna system is influenced by free-space path loss and dual phase shifts caused by signal propagation in free space and within the waveguide.

The channel model in \eqref{eqn: channel coefficient pinching} highlights the key distinctions between pinching-antenna systems and conventional antenna systems
For instance, unlike the fixed-position antenna system in \eqref{eqn: h convention}, the locations of the pinching antennas are adjustable, meaning that $\tilde x_n$s' can be optimized to enhance the channel conditions; 
while compared to the fluid/movable-antenna systems, the pinching-antenna system overcomes the limitation of wavelength-scale adjustments for antenna locations. Beyond these advantages, the phase shifts caused by signal propagation within the waveguide further distinguish the pinching-antenna system from existing flexible-antenna systems, while simultaneously introducing challenges in system design, as will be discussed later.

Then, the received signal at the user can be written as\vspace{-3mm}

\begin{small}
\begin{align}
    y^\pin = \sqrt{\frac{P}{N}}\sum_{n=1}^N h_n^\pin s + w, 
\end{align}
\end{small}%
where $s$ is the signal passed on the waveguide. 
Here, we assume that the total transmit power is evenly distributed among the $N$ active pinching antennas. Then, the received signal at the user can be written as \vspace{-3mm}

\begin{small}
   \begin{align} \label{eqn: received signal}
        y^\pin \!=\!\bigg(\!\sum\limits_{n=1}^N\! \frac{\eta^{\frac{1}{2}} e^{-j \big(\frac{2 \pi}{\lambda}\|\boldsymbol{\psi}_m - \widetilde{\boldsymbol{\psi}}_n^{\pin} \| +  \frac{2\pi}{\lambda_g}\|\widetilde{\bm{\psi}}^{\pin}_0 - \boldsymbol{\widetilde \psi}_n^{\pin}\| \big)}}{\|\boldsymbol{\psi}_m  -\tilde{\boldsymbol{\psi}}_n^{\pin} \|} \!\bigg) \sqrt{\frac{P}{N}} s + w.
    \end{align} 
\end{small}%
Using this model, the achievable data rate for the pinching-antenna system can be expressed as
\begin{small}
\begin{align}
    \!\!\!R^\pin \!=\! \log \bigg( 1 \!+\! \bigg| \sum\limits_{n=1}^{N} \!\frac{\eta^{\frac{1}{2}} e^{-j \left( \frac{2\pi}{\lambda} \| \bm{\psi}_m - \bm{\widetilde \psi}_n^{\pin} \| + \theta_n \right)}}{\| \bm{\psi}_m - \bm{\widetilde\psi}_n^{\pin} \|} \bigg|^2 \!\frac{P}{N \sigma_w^2} \bigg)\!.
\end{align}
\end{small}%
The goal of this paper is to maximize the downlink data rate by optimizing the locations of the pinching antennas. The associated optimization problem can be formulated as
\begin{subequations} \label{p: original problem}
    \begin{align}
        \max_{\tilde x_1,...,\tilde x_n} ~&R^\pin \\
        \st ~&|\tilde x_n - \tilde x_{n'}| \geq \Delta, \forall n,n' \in \Nset, \label{eqn: noncvx antenna spacing}
    \end{align}
\end{subequations}
where constraints \eqref{eqn: noncvx antenna spacing} guarantee that the antenna spacings should be no smaller than the minimum guide distance, $\Delta$, to avoid the antenna coupling.
Without loss of generality, we assume that the pinching antennas are deployed in a successive order, which means $\tilde x_n - \tilde x_{n-1} > 0, \forall n \in \Nset.$
As a result, the nonconvex constraints \eqref{eqn: noncvx antenna spacing} can be simplified as the following linear constraints:
\begin{align}
    \tilde x_n - \tilde x_{n-1} \geq \Delta, \forall n \in \Nset.
\end{align}
Meanwhile, we note that maximizing the downlink data rate is equivalent to maximizing the signal-to-noise ratio (SNR). Therefore, problem \eqref{p: original problem} can be equivalently recast as follows: \vspace{-4mm}

\begin{small}
\begin{subequations}\label{p: rate maximization}
    \begin{align}
     \max_{\tilde x_1,...,\tilde x_n} ~&\bigg| \sum\limits_{n=1}^{N} \frac{ e^{-j \phi_n }}{\| \bm{\psi}_m - \bm{\widetilde\psi}_n^{\pin} \|} \bigg|\\
     \st ~&\tilde x_n - \tilde x_{n-1} \geq \Delta, \forall n \in \Nset,\\
     ~& \phi_n = \frac{2\pi}{\lambda} \| \bm{\psi}_m - \bm{\widetilde \psi}_n^{\pin} \| + \theta_n, \forall n \in \Nset.
    \end{align}
\end{subequations}
\end{small}%
However, problem \eqref{p: rate maximization} is still challenging to solve because 
the locations of the pinching antennas affect both the numerators and denominators of the objective function, and also appear in the exponent of complex-valued numbers.
In the next section, an efficient algorithm is proposed to handle this problem.

\vspace{-3mm}
\section{Proposed Algorithm to Solve Problem \eqref{p: rate maximization}}
Looking at the objective function of the problem \eqref{p: rate maximization}, we can see that the locations of the pinching antennas affect two key aspects of the channel. 
One is the large-scale path loss, and the other is the phase shifts due to signal propagation inside and outside the waveguide. 
To maximize the objective function, we need to minimize the impact of the large scale path loss associated with the terms, $|\bm{\psi}_m - \bm{\widetilde \psi}_n^{\pin}|, \forall n \in \Nset$, on the denominator. Meanwhile, we also need to constructively combine the received signals from different pinching antennas at the user. To achieve this goal, we consider the following relaxed problem: \vspace{-3mm}

\begin{small}
\begin{subequations}\label{p: rate maximization approx 1}
    \begin{align}
     \max_{\tilde x_1,...,\tilde x_n} ~& \sum\limits_{n=1}^{N} \frac{1}{\| \bm{\psi}_m - \bm{\widetilde\psi}_n^{\pin} \|} \label{eqn: obj}\\
     \st ~&\tilde x_n - \tilde x_{n-1} \geq \Delta, \forall n \in \Nset,\\
     ~& \phi_n -\phi_{n-1} = 2k\pi, \forall n \in \Nset, \label{eqn: phase constraints}
    \end{align}
\end{subequations}
\end{small}%
where \eqref{eqn: obj} is to minimize the effects of large-scale path loss by maximizing the sum of the reciprocals of the distances, 
and the constraint \eqref{eqn: phase constraints} ensures that the received signals from different pinching antennas can be constructively combined at the user, where $k$ is an arbitrary integer. Although problem \eqref{p: rate maximization approx 1} is not strictly equivalent to problem \eqref{p: rate maximization}, our simulation results in section \ref{sec: simulation} show that solving problem \eqref{p: rate maximization approx 1} achieves almost the same performance as the optimal solution of problem \eqref{p: rate maximization}.

Another important observation is that to satisfy constraints \eqref{eqn: phase constraints}, one only needs to move the pinching antennas on the wavelength scale, which is much smaller than the distances between the pinching antennas and the user. Based on this insight, we design a low-complexity two-stage algorithm to solve the \eqref{p: rate maximization approx 1} problem. Specifically, in the first stage, we aim to maximize the sum of the reciprocals of the distances from the pinching antennas to the user under the antenna spacing constraints. Then, in the second stage, we refine the pinching antenna locations to satisfy the constraints \eqref{eqn: phase constraints} by moving them on the wavelength scale. The details of the algorithm are described in the next two subsections.

\vspace{-3mm}
\subsection{Maximize the Summation of Reciprocals of Distances}
The problem of maximizing the summation of reciprocals of distances under the antenna spacing constraints is given by \vspace{-4mm}
\begin{subequations}\label{p: rate maximization approx 2}
    \begin{align}
     \max_{\tilde x_1,...,\tilde x_n} ~& \sum\limits_{n=1}^{N} \big[(\tilde x_n - x_m)^2 + C \big]^{-\frac{1}{2}} \\
     \st ~&\tilde x_n - \tilde x_{n-1} \geq \Delta, \forall n \in \Nset, \label{eqn: antenna spacing constraints}
    \end{align}
\end{subequations}
where {\small $C = y_m^2 \!+\! d^2 > 0$}.  An interesting property of the optimal solutions of problem \eqref{p: rate maximization approx 2} is provided in the following lemma.

\vspace{-2mm}
\begin{lem} \label{lem: antenna spacing}
    With the optimal solution of problem \eqref{p: rate maximization approx 2}, the constraints \eqref{eqn: antenna spacing constraints} hold with equalities, i.e.,
    \begin{align}
        \tilde x_n^* - \tilde x_{n-1}^* = \Delta, \forall n \in \Nset,
    \end{align}
\end{lem}

{\bf Proof:} See Appendix \ref{appd: lemma antenna spacing}. \hfill $\blacksquare$

With Lemma \ref{lem: antenna spacing}, problem \eqref{p: rate maximization approx 2} can be simplified as 
\begin{align} \label{p: rate maximization approx 3}
    \max_{\tilde x_1} ~& \sum\limits_{n=1}^{N} \big[(\tilde x_1 + (n-1)\Delta - x_m)^2 + C \big]^{-\frac{1}{2}}.
\end{align}
Therefore, the original problem in \eqref{p: rate maximization approx 2} which is related to multiple optimization variables can be reduced to a simplified version with respect to the location of the first pinching antenna only.
However, problem \eqref{p: rate maximization approx 3} is still nonconvex with respect to $\tilde x_1$ and hence challenging to solve. Intriguingly, the following lemma reveals that problem \eqref{p: rate maximization approx 3} has a special structure, which paves the way to obtain the globally optimal solution of problem \eqref{p: rate maximization approx 3}.

\vspace{-2mm}
\begin{lem} \label{prop: optimal x1}
    The objective function of problem \eqref{p: rate maximization approx 3} is unimodal with respect to $\tilde x_1$ if $C \geq (N-1)^2 \Delta^2$, and the optimal $\tilde x_1$ maximizing the objective function is given by
    \begin{align}
        \tilde x_1^* = x_m - \frac{(N-1)}{2} \Delta.
    \end{align}
\end{lem}

{\bf Proof:} See Appendix \ref{appd: prop optimal x1}. \hfill $\blacksquare$
We note that the condition on $C$ is mild. For instance, when $N=8$, $f = 26$ GHz and $\Delta$ is set to half a wavelength, it only requires $C \geq 0.0306$. 
According to Lemma \ref{prop: optimal x1}, the locations of the pinching antennas can be obtained in closed-form. In particular, the location of pinching antenna $n$ is given by
\begin{align} \label{eqn: pinching antenna location}
    \widetilde {\bm\psi}_n^\pin = \Big[x_m - \big(\frac{N-1}{2} + n -1\big) \Delta, 0, d \Big], \forall n \in \Nset. 
\end{align}
It is important to emphasize that the solutions in \eqref{eqn: pinching antenna location} do not necessarily maximize the user’s received signal strength. This is because the effective channel gain is also affected by the dual phase shifts caused by signal propagation both within and outside the waveguide, as discussed under \eqref{eqn: channel coefficient pinching}. Therefore, further refinement of the pinching antenna locations is required to ensure constructive signal reception at the user.

\vspace{-4mm}
\subsection{Refine the Pinching Antenna Locations to Satisfy \eqref{eqn: phase constraints}}
The remaining of the optimization procedure is to refine the pinching antenna locations to satisfy \eqref{eqn: phase constraints}, such that signals sent by different pinching antennas are constructively combined at the user. 
The detailed descriptions are as follows, where the case that $N$ is an odd number is focused on for illustration purposes. 

\begin{itemize}
    \item According to Lemma \ref{prop: optimal x1}, the obtained location of the $(\frac{N+1}{2})$-th pinching antenna is set as $[x_m,0,d]$. 
    \item Next, the location of the  $(\frac{N+1}{2} + 1)$-th pinching antenna is refined. In particular, for computation reduction, we concentrate on the segment between $[x_m + \Delta, 0, d]$ and $[x_m + 3\Delta, 0, d]$, and the refined location is obtained by using the first-found location which minimizes ${\rm mod}\{\phi_{\frac{N+1}{2}} - \phi_{\frac{N+1}{2}+1}, 2\pi\}$, where ${\rm mod}\{a,b\}$ signifies the module operation of $a$ by $b$.
    \item Then, the location of the $n$-th ($n > \frac{N+1}{2}+1$) pinching antenna can be successively obtained by focusing on the segment between $[\tilde x_{n-1} + \Delta, 0,d]$ and $[\tilde x_{n-1} + 3\Delta, 0,d]$ by using the previously obtained location which minimizes ${\rm mod}\{\phi_{n-1} - \phi_{n}, 2\pi\}$.
    \item The locations of the pinching antennas $n < \frac{N+1}{2}$ can be found successively by following the same steps as above.
\end{itemize}
This scheme also applies to the case in which $N$ is an even number. Specifically, the location of the $\frac{N}{2}$-th pinching antenna is initially fixed based on Lemma \ref{prop: optimal x1}. Then, the locations of the remaining pinching antennas can be determined sequentially as outlined above.

\vspace{-2mm}
\section{Simulation Results} \label{sec: simulation}
In this section, the performances of the pinching-antenna systems and the proposed algorithm are evaluated via computer simulations. Without loss of generality, the same choices of the system parameters as in \cite{ding2024flexible} are used, e.g., the noise power is set to $-90$ dBm, $f_c = 28$ GHz, $d = 3$ meters, $\Delta = \frac{\lambda}{2}$, and $n_{\rm neff} = 1.4$. While the transmission power $P$, the side length ($D$) of the square area, and the number of pinching antennas $N$ are specified in each figure.

\begin{figure*}[!t]
	\begin{minipage}{0.32\linewidth}
		\centering
		\includegraphics[width=0.86\linewidth]{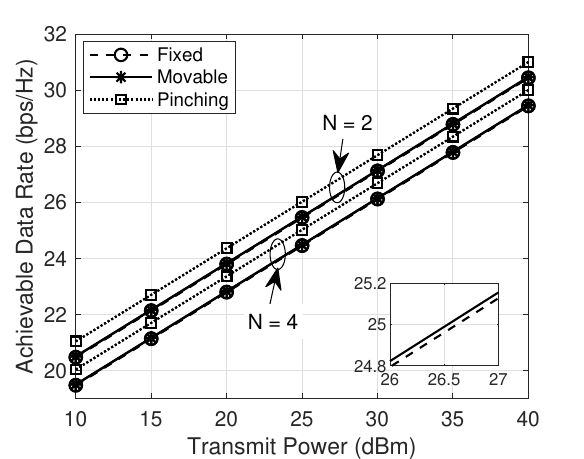}
		\captionsetup{justification=justified, singlelinecheck=false, font=small}	
	\caption{\small Data rates of the pinching-antenna and the conventional antenna systems versus transmission powers with $D = 10$ meters.} \label{fig: rate power} 
	\end{minipage}~~
	\begin{minipage}{0.32\linewidth} 
		\centering
		\includegraphics[width=0.86\linewidth]{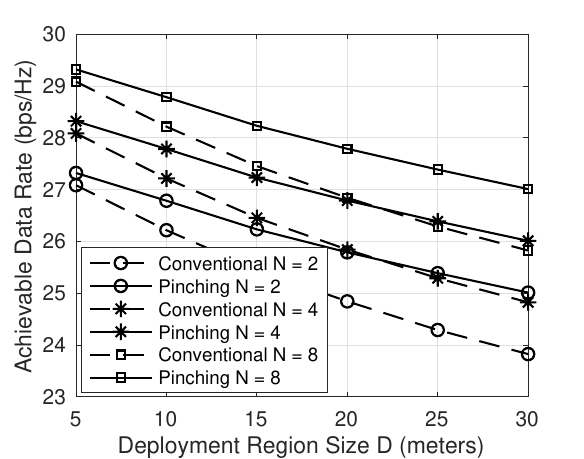}\\
        \captionsetup{justification=justified, singlelinecheck=false, font=small}
	\caption{\small Data rates of the pinching-antenna and the conventional antenna systems versus side lengths with $P = 30$ dBm.} \label{fig: rate d} 
	\end{minipage}~~
	\begin{minipage}{0.32\linewidth}
			\centering
			\includegraphics[width=0.86\linewidth]{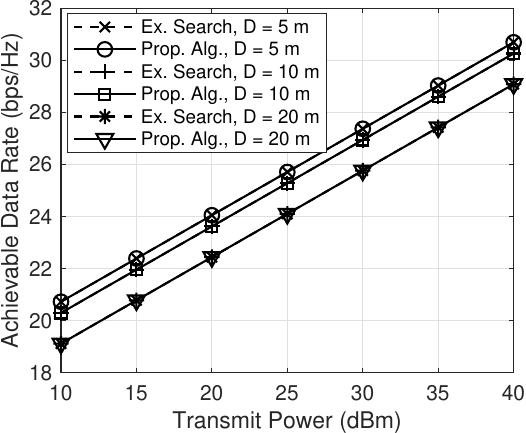}\\
        \captionsetup{justification=justified, singlelinecheck=false, font=small}	
	\caption{\small Data rates  of the proposed algorithm and the exhaustive search-based method for the pinching-antenna system with $N=2$.} \label{fig: prop search}  
	\end{minipage}
\vspace{-5mm}
\end{figure*}

Fig. \ref{fig: rate power} is provided to validate the advantages of the pinching-antenna system compared to the fixed-position antenna system and the movable-antenna system by using the ergodic achievable data rate as the performance metric. In particular, the movable antennas are assumed to be deployed on a pre-installed track parallel to the $y$-axis, with the midpoint of the track at $[0,0,d]$. The length of the track is set to $(N+9)\Delta$, allowing the maximum adjustable range for the movable antenna location to be $10 \Delta$.
As depicted in Fig. \ref{fig: rate power}, pinching-antenna systems yield higher data rates compared to conventional antenna systems.
This performance gain comes from the fact that the pinching-antenna system ensure that the pinching antennas can be flexibly deployed to the ideal locations which reduce the large-scale path loss.
Meanwhile, the achievable data rate increases with the number of antennas for both the pinching-antenna and conventional antenna systems, thanks to the additional degrees of freedom provided by using more antennas. However, it is worthy to point out that adding extra antennas in conventional antenna systems is not straightforward, whereas the pinching-antenna system offers superior flexibility to reconfigure the antenna system. 

In Fig. \ref{fig: rate d}, the performance of the pinching-antenna systems with respect to the size of the user’s deployment region is evaluated.
As can be observed from the figure, the achievable data rates of both antenna systems decrease with increasing $D$, due to larger path losses.
It is also observed that the performance gap between the pinching-antenna systems and the conventional antenna systems increases with $D$, which not only shows the great flexibility of pinching-antenna systems to compensate path losses, but also demonstrates the robustness of the pinching-antenna system to the diverse user deployment. 

To better evaluate the performance of the proposed algorithm for solving problem \eqref{p: rate maximization}, the exhaustive search method is used as a benchmark in Fig. \ref{fig: prop search}.
Specifically, the exhaustive search method solves problem \eqref{p: rate maximization} by searching all possible locations of the pinching antennas on the waveguide with a step size of $\frac{\lambda}{50}$.
Since the complexity of the exhaustive search method increases quickly with the number of pinching antennas, we focus on the special case with $N=2$ pinching antennas in Fig. \ref{fig: prop search}.
This figure shows that the proposed algorithm achieves almost the same performance as the optimal solution, confirming the efficacy of the proposed scheme, which first minimizes the effects of path loss and then refines the pinching antenna locations to ensure constructive signal combination at the user.

\vspace{-3mm}
\section{Conclusions}
In this work, we have studied the downlink rate maximization problem in a pinching-antenna system, where $N$ pinching antennas are deployed on a waveguide to serve a single-antenna user.
Since the pinching antenna locations affect both the path losses and dual phase shifts of the user’s effective channel gain, the problem is quite challenging to solve.
To address this issue, we have proposed to consider a relaxed problem and solved it with a two-stage algorithm in a low-complexity manner. 
Simulation results have demonstrated the advantages of the considered pinching-antenna system and the efficiency of proposed algorithm. 
There are several promising directions for future research. For instance, to further enhance the system’s spectral efficiency, it would be valuable to design more flexible power allocation schemes among multiple pinching antennas. Additionally, exploring system design strategies tailored for multi-user scenarios is an important avenue for making pinching-antenna systems more practical.

\vspace{-3mm}
\section*{Appendix}
\begin{appendices}
    \subsection{Proof of Lemma \ref{lem: antenna spacing}} \label{appd: lemma antenna spacing}
    Lemma \ref{lem: antenna spacing} can be proved by using contradictions. Without loss of generality, we assume that $\tilde x_n^* - \tilde x_{n-1}^* > \Delta$. 
    \begin{itemize}
        \item If $\tilde x_n^* + \tilde x_{n-1}^* \leq 2x_m$, we can move the $(n-1)$-th antenna towards the $n$-th antenna with a length $\delta_1 > 0$, such that $\tilde x_n^* - (\tilde x_{n-1}^* + \delta_1) = \Delta$ and $\tilde x_{n-1} + \delta_1 < x_m$. Meanwhile, we keep the locations of other pinching antennas unchanged. With the new location $\tilde x_{n-1} + \delta_1$, the $(n-1)$-th term of the objective function, $(\tilde x_{n-1} - x_m)^2 + C$, decreases. Therefore, the objective function value increases.
        \item If $\tilde x_n^* + \tilde x_{n-1}^* \geq 2 x_m$, we can similarly show that by moving the $n$-th antenna to location $\tilde x_n^* - \delta_2$ with $\delta_2>0$, such that $(\tilde x_n^* -\delta_2) - \tilde x_{n-1}^* = \Delta$ and $\tilde x_n^* - \delta_2 > x_m$, the objective function value increases.
    \end{itemize}
    In summary, if the obtained solutions do not make constraints \eqref{eqn: antenna spacing constraints} active, one can always find another set of locations such that constraints \eqref{eqn: antenna spacing constraints} hold with equalities and the objective function value increases. This completes the proof. \hfill $\blacksquare$

    \vspace{-2mm}
    \subsection{Proof of Lemma \ref{prop: optimal x1}} \label{appd: prop optimal x1}
    To prove Lemma \ref{prop: optimal x1}, let's first define
    \begin{small}
        \begin{align}
            \!\!\!\! g(\tilde x_1) \!\triangleq\! \sum\limits_{n=1}^{N} h_n(\tilde x_1)  \!= \!\sum\limits_{n=1}^{N}\big[(\tilde x_1 \!+\! (n\!-\!1)\Delta \!-\! x_m)^2 \!+\! C \big]^{-\frac{1}{2}}.
        \end{align}
    \end{small}
    The first-order derivative of $g(\tilde x_1)$ is given by
    \begin{small}
    \begin{align}
        g'(\tilde x_1) \!=\! \sum\limits_{n=1}^{N} h_n'(\tilde x_1) \!= \! \sum_{n=1}^N \frac{-(\tilde{x}_1 \!+\! (n\!-\!1)\Delta \!-\! x_m)}{\left[ (\tilde{x}_1 \!+\! (n\!-\!1)\Delta \!-\! x_m)^2 + C \right]^{\frac{3}{2}}}.
    \end{align}
    \end{small}%
    Next, we prove that $g(\tilde x_1)$ is unimodal with $\tilde x_1$ in three steps:
    \subsubsection {$g'(x_m - \frac{N-1}{2} \Delta) = 0$} By inserting $\tilde x_1 = x_m - \frac{N-1}{2} \Delta$ into $g'(\tilde x_1)$, we have
    \begin{small}
            \begin{align}
                \!\!\!\!g'(x_m \!-\! \frac{N\!-\!1}{2} \Delta) 
                &= \sum\limits_{n=1}^{N} \frac{\big(\frac{N+1}{2}-n\big)\Delta}{\big[\big((n-\frac{N+1}{2})\Delta\big)^2 \!+\! C\big]^{\frac{3}{2}}}. \label{eqn: first-order g}
            \end{align}
        \end{small}%
        It is straightforward to show from \eqref{eqn: first-order g} that $h_n'\big(x_m - \frac{N\!-\!1}{2}\Delta\big) = - h_{N-n+1}'\big(x_m - \frac{N\!-\!1}{2}\Delta\big)$.
        For the case that $N$ is an even number, it is clear that $g'(x_m - \frac{N-1}{2} \Delta) = 0$. For the case that $N$ is an odd number, we have $h_{\frac{N+1}{2}}'\big(x_m - \frac{N\!-\!1}{2} \Delta\big) = 0$, and thus we have $g'(x_m - \frac{N-1}{2} \Delta) = 0$.
        Summarily, $g'(x_m - \frac{N-1}{2} \Delta) = 0$ always holds.
        
        \subsubsection{$g'(\tilde x_1) > 0$ for $\tilde x_1 < x_m - \frac{N-1}{2} \Delta$, if $C \geq (N-1)^2\Delta^2$} For $\tilde x_1 < x_m - \frac{N-1}{2} \Delta$, first introduce a parameter $\delta_3 > 0$, such that $\tilde x_1 = x_m - \frac{N-1}{2} \Delta -\delta_3$. The first-order derivative of $g(\tilde x_1)$ is given by
        \begin{small}
            \begin{align}
                g'(\tilde x_1)&= \sum\limits_{n=1}^{N} h_n'\Big(x_m - \frac{N\!-\!1}{2} \Delta - \delta_3\Big)\notag\\
                &= \sum\limits_{n=1}^{N} \frac{\big(\frac{N+1}{2} - n\big)\Delta + \delta_3}{\big[\big((n -\frac{N+1}{2})\Delta - \delta_3\big)^2 \!+\! C\big]^{\frac{3}{2}}}.
            \end{align}
        \end{small}%
        Without loss of generality, we assume $n< \frac{N+1}{2}$, and we have
        \begin{small}
            \begin{align}
                &h_n'\Big(x_m \!-\! \frac{N\!\!-\!1}{2} \!\Delta - \delta_3\Big) \!=\! \frac{\big(\frac{N\!+\!1}{2} \!-\! n\big)\Delta \!+\! \delta_3}{\big[\big((\frac{N+1}{2}\!-\! n )\Delta \!+\! \delta_3\big)^2 \!+\! C\big]^{\frac{3}{2}}}, \\
                &h_{N-n+1}'\Big(\!x_m \!-\! \frac{N\!\!-\!1}{2}\! \Delta \!-\! \delta_3\!\Big) \!=\! \frac{\big(n \!-\!\frac{N+1}{2}  \big)\Delta \!+\! \delta_3}{\big[\big((\frac{N+1}{2}\!-\!n)\Delta \!-\! \delta_3\big)^2 \!+\! C\big]^{\frac{3}{2}}}.
            \end{align} 
            \end{small}%
        To let $g'(\tilde x_1) > 0$ for {\small$\tilde x_1 < x_m - \frac{N-1}{2} \Delta$}, it suffices to show {\small $h_n'\big(x_m \!-\! \frac{N\!-\!1}{2} \Delta - \delta_3\big) + h_{N-n+1}'\big(x_m \!-\! \frac{N\!-\!1}{2} \Delta - \delta_3\big)\! >\! 0, \forall n \in \Nset$}. It is noted that, since {\small$\frac{N+1}{2}\!-\!n \!>\! 0$}, we have  {\small$h_n'\big(x_m \!-\! \frac{N\!-\!1}{2} \Delta - \delta_3\big) \!>\! 0$} always hold. Meanwhile, if  {\small$\big(n \!-\!\frac{N+1}{2}  \big)\Delta \!+\! \delta_3 \!> \!0$}, we have  {\small$h_{N-n+1}'\big(x_m \!-\! \frac{N\!-\!1}{2} \Delta - \delta_3\big) \!>\! 0$}, and thus $g'(\tilde x_1) \!>\! 0$. Consequently, we only need to focus on the case that  {\small$0 \leq \delta_3 \leq \big(\frac{N+1}{2}\!-\!n\big)\Delta$}, and the condition is equivalent to {\small $\big(h_n'(x_m - \frac{N\!-\!1}{2} \Delta - \delta_3)\big)^2 \!-\! \big(h_{N-n+1}'(x_m - \frac{N\!-\!1}{2} \Delta - \delta_3)\big)^2 > 0, \forall n \in \Nset$}.
        For ease of notation, denote {\small$z \!=\! (\frac{N+1}{2}\!-\!n)\Delta \!>\! 0$}. Then, we have
        \begin{small}
        \begin{subequations}
        \begin{align}
            & \Big(h_n'\big(x_m \!-\! \frac{N\!-\!1}{2} \Delta \!-\! \delta_3\big)\Big)^2 \!-\! \Big(h_{N-n+1}'\big(x_m \!-\! \frac{N\!-\!1}{2} \Delta \!-\! \delta_3\big)\Big)^2 \notag\\
            & = \frac{(z+\delta_3)^2}{\big[(z+\delta_3)^2 + C\big]^3} - \frac{(z-\delta_3)^2}{\big[(z-\delta_3)^2 + C\big]^3} \\
            & \circeq (z+\delta_3)^2\big[(z\!-\!\delta_3)^2 \!+\! C\big]^3 \!-\! (z-\delta_3)^2 \big[(z+\delta_3)^2 \!+\! C\big]^3  \label{eqn: sign}\\
            & = 4z\delta_3 \left[ C^3 - (z^2 - \delta_3^2)^2 (2z^2 + 2\delta_3^2 + 3C) \right] \\
            & \circeq C^3 - (z^2 - \delta_3^2)^2 (2z^2 + 2\delta_3^2 + 3C) \\
            & > C^3 - (z^2 + \delta_3^2)^2 (2z^2 + 2\delta_3^2 + 3C) \label{eqn: inequality} \\
            & = C^3 - 3C(z^2 + \delta_3^2)^2 - 2 (z^2 + \delta_3^2)^3 \triangleq f(C)
        \end{align}
        \end{subequations}
        \end{small}%
        where $a \circeq b$ means $a$ and $b$ have the same sign.  By verifying the first-order derivative, it is not difficult to see that $f(C)$ is monotonically increasing with $C$ as long as 
        \begin{align}
            3C^2 - 3(z^2 + \delta_3^2)^2 > 0~ \Leftrightarrow ~ C > z^2 + \delta_3^2. \label{eqn: monotonicity}
        \end{align}
        Meanwhile, when $C = 2 (z^2 + \delta_3^2)$, we have
        \begin{align}
            f(C) & = C^3 - 3C(z^2 + \delta_3^2)^2 - 2 (z^2 + \delta_3^2)^3 \notag\\
            & = 8(z^2 \!+\! \delta_3^2)^3 \!-\! 6(z^2 \!+\! \delta_3^2)^3 \!-\! 2 (z^2 \!+\! \delta_3^2)^3 \!=\! 0. \label{eqn: positive}
        \end{align}
        Combining \eqref{eqn: monotonicity} and \eqref{eqn: positive}, we have that $f(C) \geq 0$ if $C \geq 2 (z^2 + \delta_3^2)$. 
        Since $0 \leq \delta_3 \leq z$, it suffices to set 
        \begin{align}
             C \geq 4 z^2 \geq (N-1)^2 \Delta^2.
        \end{align}
        Since $h_n'(\cdot)$ and $h_{N-n+1}'(\cdot)$ are symmetry, the same conclusion can be made to the case $n> \frac{N+1}{2}$. Therefore, we have $g'\big(x_m - \frac{N-1}{2} \Delta -\delta_3\big) > 0$ for the case that $N$ is an even number. While for the case that $N$ is an odd number, we have \vspace{-6mm}
        
        \begin{small}
        	\begin{align}
        		&h_{\frac{N+1}{2}}'\Big(x_m - \frac{N\!-\!1}{2} \Delta - \delta_3\Big)= \frac{\delta_3}{\big(\delta_3^2 \!+\! C\big)^{\frac{3}{2}}} > 0.
        	\end{align}
        \end{small}%
        Thus, $g'\big(x_m \!-\! \frac{N-1}{2} \Delta \!-\!\delta_3\big) \!>\! 0$ can be established for the case that $N$ is an odd number. 
        In summary, if $C \geq (N-1)^2 \Delta^2$, $g'(\tilde x_1) > 0$ always hold for $\tilde x_1 < x_m - \frac{N-1}{2} \Delta$. This indicates that $g(\tilde x_1)$ is monotonically increasing for $\tilde x_1 < x_m - \frac{N-1}{2} \Delta$.
        
        \subsubsection{$g'(\tilde x_1) < 0$ for $\tilde x_1 < x_m - \frac{N-1}{2} \Delta$, if $C \geq (N-1)^2\Delta^2$} Similar to step 2), we can prove that $g'(\tilde x_1)$ is strictly negative for $\tilde x_1 \!>\! x_m \!-\! \frac{N-1}{2} \Delta$ if $C \geq (N\!-\!1)^2\Delta^2$. This means that $g(\tilde x_1)$ is a monotonically decreasing function for $\tilde x_1 \!>\! x_m \!-\! \frac{N-1}{2} \Delta$.
    
    Combining the conclusions from the steps 2) and 3), we conclude that $g(\tilde x_1)$ is a unimodal function with respect to $\tilde x_1$ if $C \geq (N-1)^2\Delta^2$, and there exists a unique solution that maximizes the objective function. On the other hand, step 1) shows that $g'(x_m - \frac{N-1}{2} \Delta) = 0$, which indicates that the optimal solution of problem \eqref{p: rate maximization approx 3} is $\tilde x_1^* = x_m - \frac{N-1}{2} \Delta$.
    The proof of the lemma is complete.  \hfill $\blacksquare$
    
\end{appendices}

\vspace{-3mm}

\smaller[1]

\end{document}